\documentclass[sigconf]{acmart}

\usepackage{booktabs}
\usepackage{multirow}
\usepackage{colortbl}
\usepackage{xcolor}
\usepackage{threeparttable}
\usepackage{amsmath}
\usepackage{microtype}
\usepackage{tabularx}
\usepackage{comment}
\usepackage{rotating}
\usepackage{array}
\usepackage{graphicx}
\usepackage[font=small]{caption}

\usepackage{hyperref}

\begin{document}

\title{Analysis of Types of Inquiries in Student-AI Interaction}
\subtitle{A case study of two CS2 tasks}



\author{Matin Amoozadeh}
\affiliation{%
  \institution{University of Houston}
  \city{Houston}
  \country{USA}
}

\author{Amin Alipour}
\affiliation{%
  \institution{University of Houston}
  \city{Houston}
  \country{USA}
}

\begin{abstract}

\textbf{Background and Context.} 
Question and inquiry are integral parts of knowledge seeking and learning.
Despite their importance, students tend to not ask enough questions in the classroom. 
However, studies have shown that students interact with generative AI systems extensively in their learning and problem solving.
\textbf{Objective.} 
In this paper, we seek to better understand types of questions that students ask AI systems, how those questions evolve during problem solving and across tasks. 
\textbf{Method.} 
We use Graesser et al. taxonomy to classify students' inquiries into 18 types. We develop a few-shot learning approach to automatically classify students’ interactions with AI into those categories.
We use this system to analyze 830 interactions of CS2 students on two programming tasks.
\textbf{Findings.} 
Our results suggest that a small subset of question types accounts for the majority of student inquiries, and the types of questions students ask change substantially as the task progresses.

\noindent \textbf{Implications.}
This paper provides insights into students’ inquiry behaviors when interacting with AI, which can inform the design of AI-based instructional support and pedagogical interventions.
\end{abstract}

\maketitle
\section{Introduction}
Generative AI systems such as ChatGPT have reshaped programming education by providing students with always-available support for debugging, code generation, conceptual explanation, and problem solving. Recent research shows that computing students increasingly use generative AI systems as learning resources during programming activities \cite{kazemitabaar2024codeaid,prather2024widening}. As AI tools become increasingly integrated into programming courses, understanding how students interact with these systems has become an important challenge for computing education research \cite{amoozadeh2024student}.

Question asking plays a central role in learning, help-seeking, and self-regulated problem solving. Prior educational research has shown that the quality and type of students’ questions reflect underlying cognitive processes, conceptual understanding, and metacognitive regulation \cite{graesser1994question}. In computing education, help-seeking behaviors are strongly associated with students’ ability to overcome programming difficulties and engage productively with learning resources \cite{marwan2020unproductive}. Generative AI changes this help-seeking landscape by enabling students to engage in continuous conversational interactions during programming tasks. Rather than asking isolated questions to instructors or peers, students can iteratively refine prompts, request explanations, validate solutions, and delegate subtasks to AI assistants in real time.

Although recent studies have examined students’ perceptions, usage patterns, help-seeking preferences, and experiences with generative AI in programming courses, much of the existing literature focuses on overall usage, perceptions, and learning or problem-solving outcomes \cite{hou2024help,scholl2024novice,prather2024widening}. Comparatively little is known about the structure and evolution of students’ questions during AI-supported programming problem solving. Existing work has not sufficiently examined how students transition between different forms of inquiry over time, how questioning behavior changes across programming tasks, or whether AI interactions reflect deeper conceptual engagement versus surface-level verification behavior. Understanding these interaction patterns is important because they may reveal whether students use AI as a collaborative learning partner or primarily as an answer-checking tool.

To analyze these interactions, we adapted the question taxonomy proposed by Graesser et al. \cite{graesser1994question} to categorize the different types of inquiries students make during AI-supported programming tasks. The taxonomy classifies questions according to their cognitive and functional roles, including verification, procedural guidance, causal reasoning, interpretation, and evaluation. Originally developed to study question asking in tutoring and educational dialogs, the framework enables fine-grained analysis of learners’ informational intent. In the context of programming education, adapting this taxonomy allows us to examine how students use AI assistants for debugging, conceptual understanding, implementation guidance, and solution validation.

In this paper, we analyze 830 prompts collected from CS2 students interacting with an AI-supported programming environment during two laboratory sessions focused on object-oriented programming in C++. Using a few-shot learning approach grounded in the Graesser taxonomy, we examine the distribution of student question types, transitions between inquiry states, and changes in questioning behavior across sessions. Our findings reveal that students frequently rely on assertion and verification prompts during early interactions, while later sessions exhibit greater use of procedural and causal reasoning questions. We further identify recurring verification loops in which students repeatedly use AI for confirmation rather than conceptual exploration.

Guided by this goal, we address the following research questions:

\textbf{RQ1:} What types of questions do students ask when interacting with an AI during programming tasks?

\textbf{RQ2:} How do students transition between different types of questions during their interaction with AI?

\textbf{RQ3:} How do students’ questioning behaviors change across programming tasks?

Our results suggest that students’ question types shift substantially as tasks progress, differ across conceptual areas, and reflect varying degrees of cognitive engagement. For example, students tend to use assertion-type questions to confirm understanding of introductory concepts such as access modifiers in object-oriented programming, while increasingly relying on AI to make program design decisions for more misconception-prone topics such as polymorphism in inheritance. 
These findings provide insights for future AI-supported instructional systems in computing education.
This paper:
\begin{itemize}
    \item applies the Graesser et al. taxonomy to student-AI programming interactions;
    \item develops a few-shot approach for inquiry classification;
    \item analyzes how inquiry patterns evolve across two CS2 programming tasks.
\end{itemize}

\section{Related Work}
\paragraph{Help-Seeking in Computing Education}
Help-seeking is an essential component of self-regulated learning and programming problem solving.
The emergence of generative AI introduces a fundamentally different form of help-seeking interaction because students can engage in continuous conversational exchanges with AI assistants while programming. Recent work by Hou et al. examined how generative AI influences computing students’ help-seeking preferences and found that AI systems are increasingly becoming part of students’ help-seeking ecosystems \cite{hou2024help}. However, their findings also suggest that effective AI-supported help-seeking requires new skills related to prompt construction and interaction strategies. Other studies similarly report that students vary considerably in how they leverage AI tools during programming activities \cite{zastudil2023generative}. While prior work has examined students’ perceptions of AI tools and help-seeking preferences, much of the literature focuses on surveys, interviews, or learning outcomes rather than the structure of students’ conversational interactions with AI systems during authentic programming tasks.

\paragraph{Question Asking and Educational Dialogue}
Question asking plays a central role in learning, tutoring, and knowledge construction. Graesser and Person proposed a widely used taxonomy of question categories that characterizes questions according to their cognitive and functional roles, including verification, causal reasoning, procedural inquiry, interpretation, and evaluation \cite{graesser1994question}. This framework has been extensively applied in educational dialogue research and intelligent tutoring systems to analyze how learners seek information and construct understanding during problem solving.

Educational dialogue research has demonstrated that the types of questions students ask often reflect underlying cognitive processes, misconceptions, and levels of engagement \cite{graesser1994question}. Prior studies in intelligent tutoring systems have shown that analyzing inquiry behavior can provide insights into productive and unproductive learning strategies during tutoring interactions. More recent work on conversational AI in education similarly emphasizes the importance of pedagogically informed dialogue systems that employ appropriate instructional strategies rather than simply generating answers \cite{kwon2024biped}. 

Despite extensive use of question taxonomies in educational dialogue research, relatively little work has adapted these frameworks to analyze conversational interactions between students and generative AI systems in programming education contexts.

\paragraph{Student-AI Interactions and Prompting Behavior}
As conversational AI systems become increasingly common in programming education, researchers have begun investigating how students construct prompts and interact with AI assistants during programming tasks. Prior studies have examined how students use generative AI during programming tasks for purposes such as conceptual understanding, code generation, and debugging \cite{scholl2024novice}. Research has also explored how students iteratively refine prompts and use AI systems for implementation support, debugging assistance, and code validation \cite{phung2023generative}. 

However, most existing work focuses primarily on performance outcomes, perceptions, or isolated prompt characteristics rather than the evolution of inquiry behavior across interactions. In particular, little research has examined how students transition between different forms of questions during AI-supported programming problem solving. Existing studies rarely combine educational question taxonomies with sequential interaction analysis to characterize how inquiry strategies evolve over time.
To address this gap, our work adapts the Graesser question taxonomy to AI-supported programming interactions and analyzes how CS2 students transition between different forms of inquiry while interacting with an AI assistant during programming tasks.

\section{Methodology}
\subsection{Study Setting and Participants}

\begin{table}[t]
\caption{Participant Characteristics by Session}
\label{tab:participants}
\centering

\small
\setlength{\tabcolsep}{3.5pt}
\renewcommand{\arraystretch}{1.08}

\begin{tabular}{lrr}
\toprule
\textbf{Characteristic} & \textbf{Session 1} & \textbf{Session 2} \\
                        & \textbf{n (\%)}    & \textbf{n (\%)}    \\
\midrule

Freshman           & 24 (40.0) & 17 (45.9) \\
Sophomore          & 21 (35.0) & 12 (32.4) \\
Junior             & 13 (21.7) &  6 (16.2) \\
Senior             &  2 (3.3)  &  2 (5.4)  \\

\midrule

Computer Science      & 44 (73.3) & 32 (86.5) \\
Computer Engineering  &  8 (13.3) &  1 (2.7)  \\
Mathematics           &  3 (5.0)  &  1 (2.7)  \\
Other Majors          &  5 (8.3)  &  3 (8.1)  \\

\midrule

Male               & 46 (76.7) & 26 (70.3) \\
Female             & 13 (21.7) & 10 (27.0) \\
Prefer not to say  &  1 (1.7)  &  1 (2.7)  \\

\midrule

First-Generation      & 26 (43.3) & 20 (54.1) \\
Continuing-Generation & 34 (56.7) & 17 (45.9) \\

\midrule

Programming Exp. (yrs) 
& $M=2.25$, $SD=1.65$
& $M=2.59$, $SD=1.71$ \\

\bottomrule
\multicolumn{3}{l}{\footnotesize \textit{N}=60 for Session 1; \textit{N}=37 for Session 2.}
\end{tabular}
\end{table}

This study was conducted during the Fall 2025 semester in a second-year
computer science course (CS2) at a large public university in the United
States. The course focuses on object-oriented programming concepts using
C++. Data were collected during scheduled laboratory sessions in which
students worked on programming assignments under the supervision of
teaching assistants.

The study included two data collection sessions conducted during separate laboratory meetings during the Fall 2025 semester, each lasting approximately 60 minutes. Session 1 was conducted during the middle of the semester, and Session 2 was conducted two weeks later. The two programming tasks differed in both topic and complexity, allowing us to examine how students’ interactions with the AI assistant varied across different object-oriented programming concepts.

Session~1 focused on foundational object-oriented programming concepts including constructors, encapsulation, and method implementation in C++. Session~2 focused on inheritance and polymorphism using related \texttt{Pet} and \texttt{Dog} classes. The second task involved more advanced object-oriented programming concepts and greater conceptual complexity.
Because the two sessions involved different programming topics and conceptual complexity, differences in students’ questioning behavior may reflect both evolving interaction patterns with the AI assistant and differences in task demands.

A total of 72 unique students participated across the two sessions. Session 1
included 60 students, while Session 2 included 37 students. Twenty-five
students participated in both sessions, allowing us to observe how
students' interactions with the AI assistant evolved across different
points in the semester.

During the lab sessions, students worked individually on their programming
assignments using a web-based programming environment with integrated AI
support developed for this study. The system logged students' prompts and
interactions with the AI assistant while they worked on their tasks.

Participation occurred as part of a graded laboratory activity within the
course, and all interactions took place in the regular classroom
environment with teaching assistants present. The study protocol was
approved by the university's Institutional Review Board (IRB). Table~\ref{tab:participants} summarizes the demographic characteristics
of the participants in each session. Most participants were freshmen or
sophomores, reflecting the typical enrollment profile of CS2 courses.

For this study, we developed a web-based programming environment that allowed students to interact with an AI assistant through a chat interface. The system integrates the programming assignment description, a code editor, program execution output, and a conversational AI assistant within one single interface. The AI assistant was powered by a large language model (GPT-4) accessed through the OpenAI API.

Upon logging in, participants first completed a short demographic survey and were then directed to the programming environment where they worked on their assignment. The interface enabled students to read the assignment instructions, write and modify code, run their program, and interact with the AI assistant through the chat interface. Students could freely ask questions and include code snippets in their prompts while receiving AI-generated guidance related to debugging, implementation strategies, and conceptual understanding of object-oriented programming in C++.

All interactions with the system were automatically logged, including student prompts, AI responses, timestamps, user identifiers, and additional interaction metadata. These logs enable detailed analysis of students’ question-asking behavior during programming tasks. The system was used during scheduled laboratory sessions lasting approximately 60 minutes. Students were allowed unlimited interaction with the AI assistant during the session. At the end of the session, students were required to submit their programming assignment through the system; if a submission was not completed manually, the system automatically submitted the current code when the time limit expired.

\begin{table*}[t]
\centering
\caption{Graesser et al \cite {graesser1994question} Question Categories}
\label{tab:graesser_categories}
\begin{tabularx}{\textwidth}{l X X}
\toprule
\textbf{Category} & \textbf{Definition} & \textbf{Example (Programming Context)} \\
\midrule
Verification & Asks whether a statement or condition is true or false & Is this constructor required? \\

Disjunctive & Asks which of multiple alternatives is correct & Should this be public or protected? \\

Concept Completion & Requests missing information needed to complete a concept & What is the return type of this function? \\

Feature Specification & Asks about properties, attributes, or components of an entity & What data members should this class have? \\

Quantification & Requests numerical or quantitative information & How many parameters does this function take? \\

Definition & Asks for the meaning of a concept or term & What does \texttt{protected} mean? \\

Example & Requests an illustrative instance of a concept & Can you give an example of inheritance? \\

Comparison & Asks about similarities or differences between concepts & What is the difference between \texttt{private} and \texttt{protected}? \\

Interpretation & Asks for explanation or inference about observed behavior or data & What is happening in this output? \\

Causal Antecedent & Asks why an event or state occurred & Why is this line causing an error? \\

Causal Consequence & Asks what will happen as a result of an action & What happens if I remove this constructor? \\

Goal Orientation & Asks about the purpose or intention behind an action & What is the goal of this function? \\

Enablement & Asks what allows an action or process to occur & What enables polymorphism here? \\

Instrumental / Procedural & Asks how to perform an action or procedure & How do I initialize the constructor? \\

Expectation & Asks what outcome should normally occur & What should the output look like? \\

Judgment / Evaluation & Evaluates correctness, quality, or appropriateness & Is this implementation correct? \\

Assertion & Declarative statement indicating belief, confusion, or knowledge & I don’t understand inheritance. \\

Request / Directive & Command requesting the listener to perform an action & Fix this code. \\
\bottomrule
\end{tabularx}
\end{table*}

\subsection{Classification using Few-Shot Learning}

Student prompts were categorized using the question taxonomy proposed by Graesser et al.~\cite{graesser1994question}. This taxonomy classifies student questions into 18 categories that reflect different cognitive functions of inquiry, including verification, explanation, causal reasoning, procedural guidance, and evaluation. Originally developed to study question asking in tutoring and instructional dialog, the framework has been widely used in educational research to analyze how learners seek information and construct understanding during problem solving.

In the context of programming tasks, adapting this taxonomy enables fine-grained analysis of how students use AI assistants for conceptual understanding, debugging support, implementation guidance, and solution validation. For example, the taxonomy allows us to distinguish between prompts seeking conceptual clarification (e.g., definition or explanation questions), procedural guidance (e.g., how to implement a feature), debugging explanations (e.g., causal antecedent questions), and confirmation of correctness (e.g., verification or judgment questions). Such distinctions provide insight into the different inquiry strategies students employ while interacting with AI during programming activities. Table~\ref{tab:graesser_categories} summarizes the question categories used in our analysis along with definitions and examples adapted to programming contexts.

To automatically classify student prompts into the Graesser categories, we employed a few-shot learning approach using large language models accessed through the OpenAI and Anthropic APIs. Few-shot learning enables the model to perform classification by conditioning on a small set of labeled examples rather than requiring task-specific supervised training. For each prompt, the models were provided with the definitions of the 18 Graesser categories together with several example prompts illustrating each category in a programming context. The models were instructed to assign each student prompt to the single most appropriate category; prompts that did not clearly fit any category were labeled as \textit{Other}.

Before classification, the dataset was preprocessed to remove empty or non-informative prompts (e.g., greetings such as ``Hi''). The remaining prompts were then submitted individually to both GPT-5.2 and Claude using a predefined classification template. To improve reliability, two authors independently reviewed the categorized prompts and compared the classifications generated by the two language models. Disagreements between models or reviewers were manually examined and resolved through discussion based on the definitions provided in the Graesser taxonomy. This validation process helped ensure that the assigned categories aligned with the intended cognitive functions of each question type. The resulting classifications were subsequently used to analyze patterns in students’ question-asking behavior during AI-supported programming tasks.

\subsection{Data Analysis}

Following the few-shot classification of prompts into the Graesser question categories, we conducted a quantitative descriptive analysis of students’ inquiry behavior during programming tasks. In total, 830 prompts were analyzed, including 432 prompts from Session~1 and 398 prompts from Session~2.

First, we examined the overall distribution of prompts across the Graesser categories to identify the most common types of questions students asked when interacting with the AI assistant. We also reported the distribution of question categories for all participants and separately for first-generation and continuing-generation students to provide descriptive insights into students’ inquiry patterns.

Next, we compared the distribution of question categories between the two sessions to explore how questioning behaviors changed over time. Because 25 students participated in both sessions, we also examined how inquiry patterns evolved for these common participants across sessions.

Finally, we analyzed the sequence of question types within each session using a state-machine representation of the Graesser categories. This analysis allowed us to examine transitions between question types and to explore how students’ inquiry strategies evolved during the progression of a programming task.

\section{Results}
\subsection{Frequency of Inquiry Types}

Before analyzing the types of questions students asked, we first summarize the dataset used in this study. Table~\ref{tab:dataset_generation} provides an overview of the question dataset collected across the two laboratory sessions. In total, 830 prompts were recorded from 97 session participations. Session~1 included 60 students who generated 432 inquiries (an average of 7.20 prompts per student), while Session~2 involved 37 students who produced 398 inquiries (an average of 10.76 prompts per student). The higher average number of prompts in Session~2 suggests that students interacted more frequently with the AI assistant during the later session.

\begin{figure}[t]
\centering
\includegraphics[width=\columnwidth]{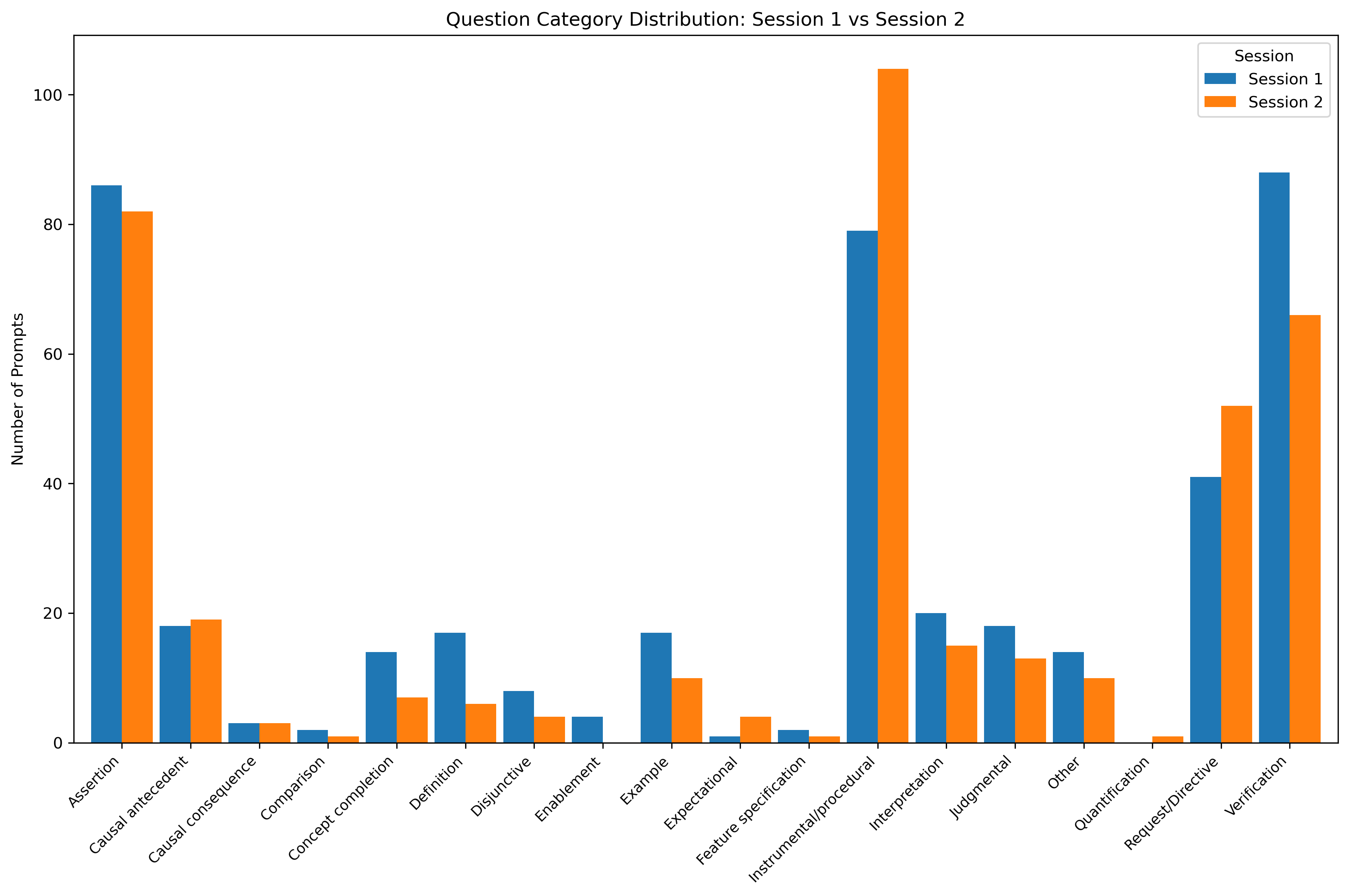}
\caption{Frequency of types of questions in Session~1 and Session~2.}
\label{fig:session_comparison}
\end{figure}

Figure~\ref{fig:session_comparison} presents the distribution of question categories across Session~1 and Session~2, based on the Graesser et al. question taxonomy. Assertion—defined as statements of confusion, lack of understanding, or problem reports without a specific question—was the most frequent category in both sessions (S1 = 86, S2 = 82), reflecting students’ tendency to report issues or express uncertainty rather than formulate explicit questions. In contrast, Instrumental/procedural prompts became the dominant category in Session~2 (S2 = 104; S1 = 79), indicating that students increasingly asked how to accomplish tasks or follow specific steps as they engaged with more complex programming challenges.

Verification prompts, which seek yes/no confirmation of correctness, decreased from Session~1 to Session~2 (S1 = 88, S2 = 66), while Request/Directive prompts—where students directly asked the assistant to perform an action—increased (S1 = 41, S2 = 52). Additionally, categories such as Interpretation and Judgment showed moderate presence in both sessions but remained secondary compared to dominant categories. Higher-order categories such as Comparison, Expectational, Feature specification, and Enablement remained minimal across both sessions, suggesting that students rarely engaged in comparative, predictive, or deeply inferential questioning.

A chi-square test of independence revealed a near-significant difference between the two sessions, $\chi^2(17) = 27.03$, $p = .0576$, indicating a trend toward a shift in questioning behavior from confusion-reporting in Session~1 toward more procedural and task-oriented inquiry in Session~2, though this difference did not reach conventional statistical significance.

\begin{table}[t]
\caption{Question Dataset Summary by Session and Generation Status}
\label{tab:dataset_generation}
\centering

\small
\setlength{\tabcolsep}{5pt}
\renewcommand{\arraystretch}{1.1}

\begin{tabular}{lcccc}
\toprule
& \multicolumn{2}{c}{\textbf{Session 1}} 
& \multicolumn{2}{c}{\textbf{Session 2}} \\
\cmidrule(lr){2-3} \cmidrule(lr){4-5}

\textbf{Category} 
& \textbf{$N$} 
& \textbf{Questions} 
& \textbf{$N$} 
& \textbf{Questions} \\
\midrule

All Students      & 60 & 432 & 37 & 398 \\
Continuing-Gen.   & 34 & 271 & 17 & 239 \\
First-Gen.        & 26 & 161 & 20 & 159 \\

\midrule

\multicolumn{5}{l}{\textit{Avg. Questions per Student (Mean $\pm$ SD)}} \\

All Students      &  & 7.20 &  & 10.76 \\
Continuing-Gen.   &  & 7.97 (7.23) &  & 14.06 (17.78) \\
First-Gen.        &  & 6.19 (5.58) &  & 7.95 (7.76) \\

\bottomrule
\end{tabular}

\smallskip
\footnotesize
\textit{Note.} $N$ = number of students.
\end{table}

Although the number of participants in the two groups is relatively comparable, first-generation students asked fewer questions overall than continuing-generation students (Table~\ref{tab:dataset_generation}). For example, in Session~1, 26 first-generation students asked 161 questions, whereas 34 continuing-generation students asked 271 questions. A similar pattern is observed in Session~2, where 20 first-generation students asked 159 questions compared to 239 questions asked by 17 continuing-generation students.

In Session~1, first-generation students relied most on 
\textit{Verification} ($\approx$26\%) and \textit{Assertion} 
($\approx$19\%), while continuing-generation students favored 
\textit{Instrumental/Procedural} ($\approx$22\%) and \textit{Assertion} 
($\approx$21\%). Between sessions, first-generation students showed a 
marked increase in \textit{Instrumental/Procedural} questions 
($\approx$12.5\% to $\approx$27\%) and declines in 
\textit{Interpretation} and \textit{Request/Directive}, whereas 
continuing-generation students shifted toward \textit{Verification} 
($\approx$8.5\% to $\approx$18\%) and \textit{Request/Directive} 
questions. These patterns are descriptive; no shifts reached statistical 
significance after Bonferroni correction.

\subsection{Inquiry Patterns by Generation}

\begin{figure}[t]
\centering
\includegraphics[width=\linewidth]{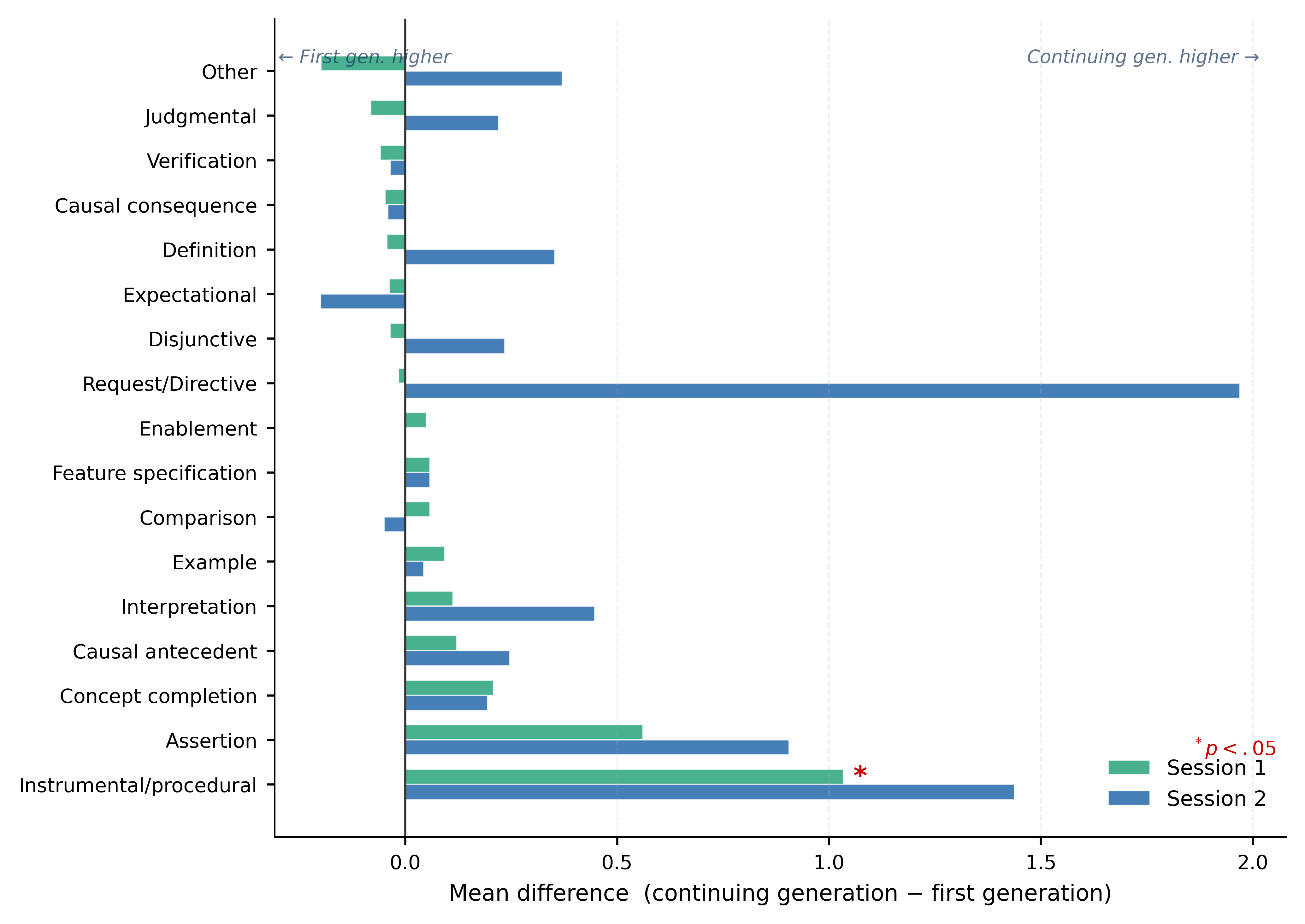}
\caption{Mean difference in Graesser question-type usage between continuing-generation and first-generation students across both sessions. Positive values indicate higher usage by continuing-generation students. An asterisk (*) denotes statistical significance at $p < .05$.}
\label{fig:pertype}
\end{figure}

Beyond aggregate diversity scores, a granular examination of individual
Graesser question-type usage reveals meaningful differences in how
first-generation and continuing-generation students engage with the AI
chatbot. Figure~\ref{fig:pertype} presents the mean difference in
usage for each of the 17 question types across both sessions, with
positive values indicating higher usage by continuing-generation
students. Across both sessions, continuing-generation students
demonstrated higher mean usage in the majority of question types
(Session~1: 9 of 17; Session~2: 12 of 17), with the most pronounced
advantages concentrated in procedural and directive categories.
Specifically, \textit{Instrumental/procedural} questions — in which
students ask the AI \textit{how} to accomplish a task — showed the
largest statistically significant difference in Session~1
($\Delta = +1.03$, $p = .013$), and remained the second-largest
advantage in Session~2 ($\Delta = +1.44$). Similarly,
\textit{Request/Directive} questions, in which students issue direct
commands or requests to the AI, favoured continuing-generation students
in both sessions ($\Delta = -0.02$ in Session~1;
$\Delta = +1.97$ in Session~2). By contrast, first-generation students
demonstrated relatively higher usage of \textit{Verification} questions
— asking whether a given answer or piece of code is correct — and
\textit{Expectational} questions, which explore hypothetical outcomes.
This pattern suggests that continuing-generation students approach the
AI as an active problem-solving partner, directing it toward task
completion, while first-generation students tend to adopt a more
confirmatory role, using the AI to check and validate rather than
to generate or explore. This distinction, replicated across both
independent sessions, constitutes the primary qualitative finding of
this study.

\begin{figure}[t]
\centering
\includegraphics[width=\linewidth]{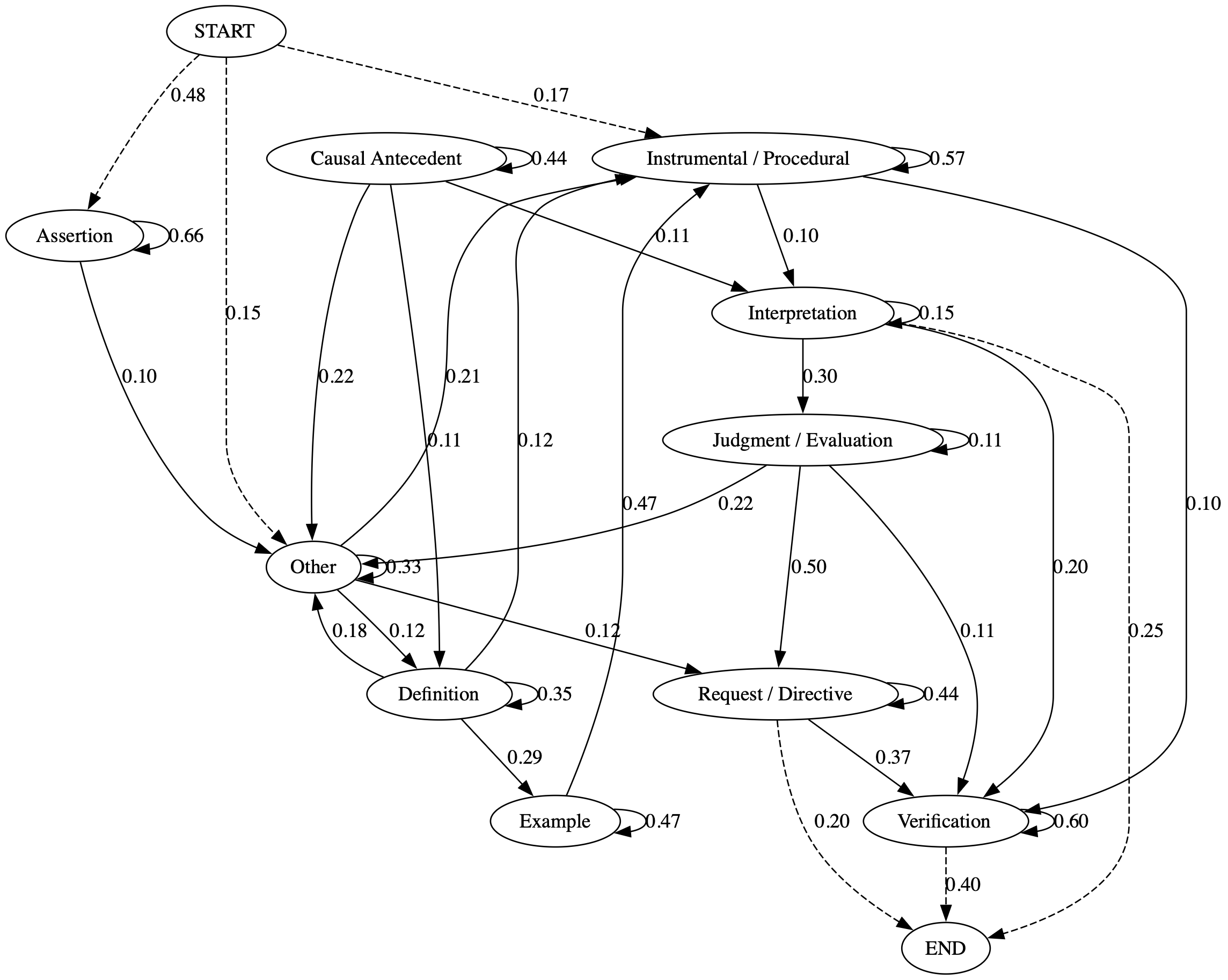}
\caption{State-transition diagram of Graesser question types in Session~1 ($N=60$ students, 432 prompts).}
\label{fig:graesser_state1}
\end{figure}

\begin{figure}[t]
\centering
\includegraphics[width=\linewidth]{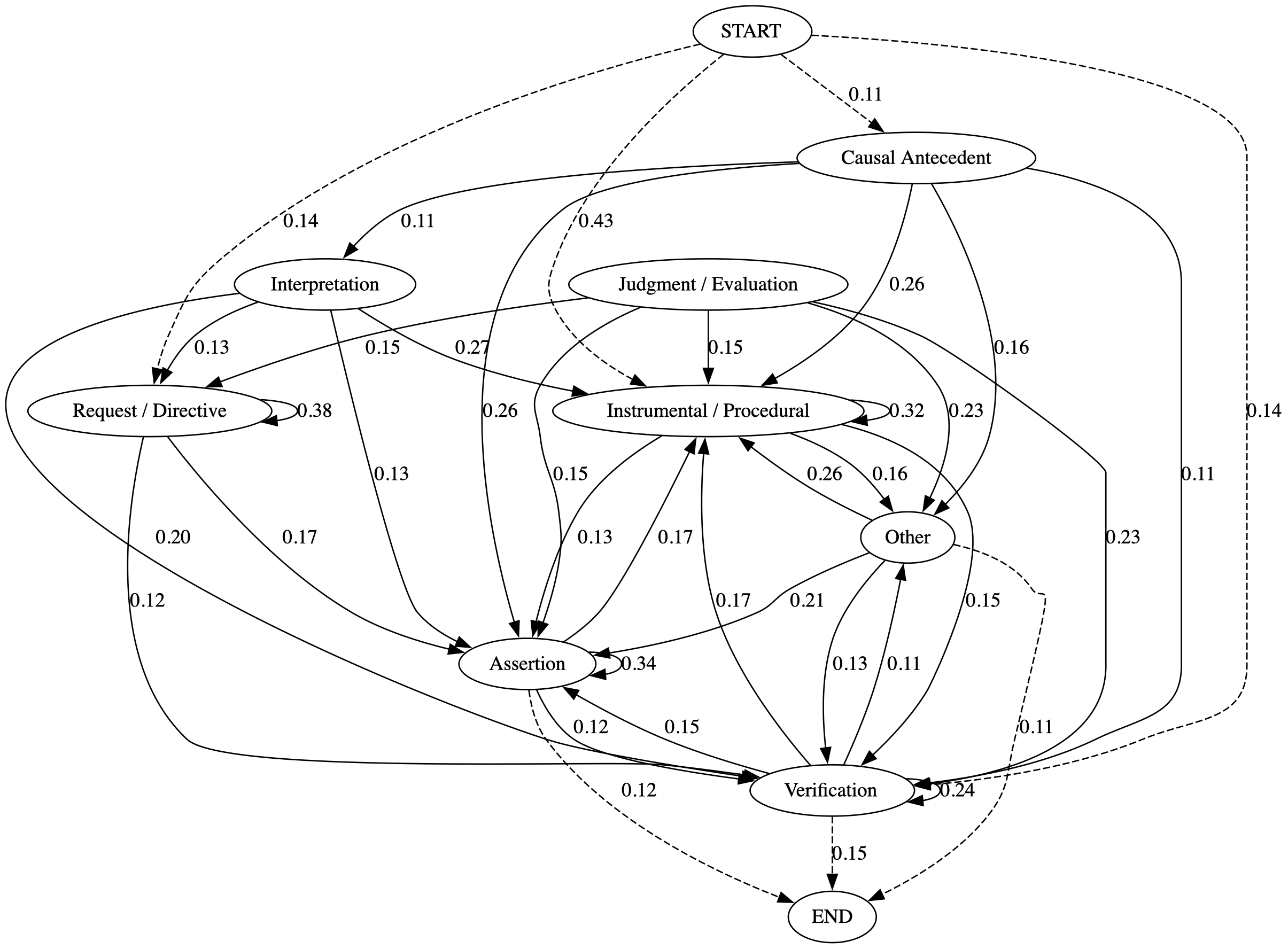}
\caption{State-transition diagram of Graesser question types in Session~2 ($N=37$ students, 398 prompts).}
\vspace{-5mm}
\label{fig:graesser_state2}
\end{figure}

 The two state machine diagrams Figures~\ref{fig:graesser_state1} and \ref{fig:graesser_state2} depict how type of students' questions changes in each session. 
 In session 1, most students start with Assertion questions (prob=0.48).  In session 2, students used a wider variety of question types.  They mostly start with an Instrumental/Procedural question (prob=0.43).
 We note that the second session took place  two weeks after the first session in which  the students gained more maturity in their questions-asking skills. 
 In both sessions, students tended to remain longer within Assertion, Request/Directive, and Instrumental/Procedural question states.

\section{Discussion and Concluding Remarks}

\paragraph{First-Generation Students and AI-Supported Help-Seeking}
We observed that first-generation students asked fewer questions than continuing-generation students. Prior research has identified differences in academic experiences and engagement associated with first-generation status~\cite{stephens2012cultural,stephens2012unseen}. Although conversational AI may reduce some interpersonal barriers associated with classroom help-seeking, differences in inquiry behavior persisted in our study. In particular, during the second task involving more complex object-oriented programming concepts, first-generation students asked substantially fewer questions. These findings suggest that AI-supported tutoring systems may benefit from scaffolding mechanisms that encourage students to formulate questions and engage in more active help-seeking behaviors.

\paragraph{Dynamics of Inquiry in AI-Supported Programming}
A central theme emerging from our findings is that students’ questioning behaviors evolved as programming tasks became more conceptually demanding. In Session 1, students relied more heavily on assertion and verification prompts, often seeking confirmation of understanding for foundational concepts. Such behavior is consistent with prior work showing that novice programmers may engage in ineffective help-seeking behaviors, including seeking help too frequently or avoiding help when it is needed~\cite{marwan2020unproductive}. In contrast, Session 2 showed increased use of instrumental/procedural and request/directive prompts, suggesting a shift toward more directive and procedural forms of AI usage. This difference may reflect increasing familiarity with the conversational AI system, differences in task complexity, or evolving strategies for interacting with AI during programming activities. Importantly, the second assignment focused on inheritance and polymorphism, concepts for which learners have been shown to experience persistent difficulties and misconceptions in object-oriented programming~\cite{liberman2011difficulties}. Additionally, the observation that a small subset of question types accounted for the majority of interactions suggests that student-AI conversations are often dominated by repeated help-seeking patterns. These findings highlight the importance of designing AI-supported programming environments that encourage reflective inquiry and productive help-seeking rather than repeated confirmation-seeking or excessive procedural delegation. However, because the two sessions differed in both programming topics and conceptual complexity, these findings should be interpreted cautiously.

\bibliographystyle{ACM-Reference-Format}
\bibliography{bib.bib}

\end{document}